\documentclass{ifacconf}
\usepackage{graphicx}      
\usepackage{natbib}        
\usepackage{amsmath}
\usepackage{units}
\usepackage{amssymb}
\usepackage{dsfont}
\usepackage{comment}
\usepackage{booktabs}
\usepackage{marginnote}
\usepackage[dvipsnames]{xcolor}
\newcommand{\flexbrac}[1]{\if\relax\detokenize{#1}\relax \else (#1) \fi}
\newcommand{\flexcomma}[1]{\if\relax\detokenize{#1}\relax \else ,#1 \fi}

\newcommand{\sN}{\mathbb{N}}

\newif\ifmargincomments 
\margincommentstrue

\ifmargincomments

\else

\fi
\begin{document}
\begin{frontmatter}

\title{Electric Autonomous Mobility-on-Demand: Joint Optimization of Routing and Charging Infrastructure Siting
\thanksref{footnoteinfo}} 

\thanks[footnoteinfo]{This publication is part of the project NEON with project number 17628 of the research program Crossover which is (partly) financed by the Dutch Research Council (NWO). }

\author[First]{F. Paparella}, 
\author[Second]{K. Chauhan}, 
\author[First]{T. Hofman},
\author[First]{M. Salazar}

\address[First]{Control Systems Techonology section, Eindhoven University of Technology, 5600 MB Eindhoven, The Netherlands.\\
e-mail: \{f.paparella,t.hofman,m.r.u.salazar\}@tue.nl.}
\address[Second]{e-mail: k.s.chauhan@student.tue.nl.}

\begin{abstract}                
The advent of vehicle autonomy, connectivity and electric powertrains is expected to enable the deployment of Autonomous Mobility-on-Demand systems.
Crucially, the routing and charging activities of these fleets are impacted by the design of the individual vehicles and the surrounding charging infrastructure which, in turn, should be designed to account for the intended fleet operation.
This paper presents a modeling and optimization framework 
where we optimize the activities of the fleet jointly with the placement of the charging infrastructure.
We adopt a mesoscopic planning perspective and devise a time-invariant model of the fleet activities in terms of routes and charging patterns, explicitly capturing the state of charge of the 
vehicles by resampling the road network as a digraph with iso-energy arcs.
Then, 
we cast the problem as a mixed-integer linear program that 
guarantees global optimality and can be solved in less than \unit[10]{min}.
Finally, we showcase 
two case studies with real-world taxi data in Manhattan, NYC: The first one captures the optimal trade-off between charging infrastructure prevalence and the empty-mileage driven by the fleet. We observe that jointly optimizing the infrastructure siting significantly outperforms heuristic placement policies, and that increasing the number of stations is beneficial only up to a certain point.
The second case focuses on vehicle design and shows that deploying vehicles equipped with a smaller battery results in the lowest energy consumption: Although necessitating more trips to the charging stations, such fleets require about $12$\% less energy than the vehicles with a larger battery capacity.

\end{abstract}
\begin{keyword}
Autonomous mobility-on-demand, electric vehicles, transportation systems.
\end{keyword}
\end{frontmatter}


\section{Introduction}\label{sec:Intro}
The electrification of the automotive industry along with the development of autonomous driving technology will lead to a paradigm shift in urban mobility.
Autonomous driving electric vehicle fleets are starting to be deployed worldwide to provide Electric Autonomous Mobility-on-Demand services (E-AMoD).
The routing and charging activities are both strongly influenced by the single vehicle design and the available charging infrastructure. Thus, all these terms play a key role in the optimization of the performance of an operational E-AMoD fleet.
Against this background, this paper proposes a modeling and optimization framework to optimize the control of an electric AMoD system jointly with the siting of the charging infrastructure, while comparing different vehicles to minimize the overall vehicle flow.
\begin{figure}[t]
	\centering
	\includegraphics[width=7cm]{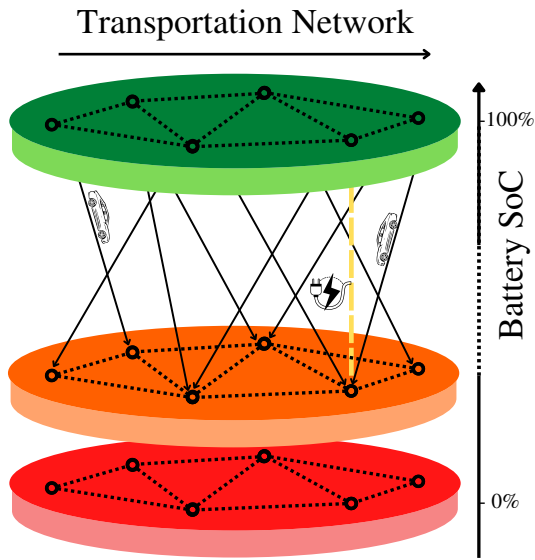}    
	\caption{Multi-layer digraph schematically representing an E-AMoD system. Each layer corresponds to a battery state-of-charge (SoC). Each node on the same vertical line represents the same geographic location. The yellow arc indicates the presence of a charging station.}
	\label{fig:Overview}
\end{figure}

\textit{Related Literature:} 
This paper pertains to the research streams of routing and charge scheduling, charging infrastructure siting, and system-level design of vehicles for an AMoD system. 
Multiple approaches to characterize and control AMoD systems are available: From queuing-theoretical models  ~\citep{ZhangPavone2016,BanerjeeJohariEtAl2015,IglesiasRossiEtAl2017}, to simulation-based ones ~\citep{LevinKockelmanEtAl2017,MaciejewskiBischoffEtAl2017,HorlRuchEtAl2019}. Multi-commodity network flow models,~\citep{RossiZhangEtAl2017,SpieserTreleavenEtAl2014,IglesiasRossiEtAl2018, SalazarLanzettiEtAl2019} are suited for efficient optimization and allow for the implementation of a variety of complex constraints. They have been successfully employed to minimize fleet travel and electricity costs subject to limited driving range, charging constraints imposed by the congestion on the power transmission grid~\citep{RossiIglesiasEtAl2018b,EstandiaSchifferEtAl2019,BoewingSchifferEtAl2020}.
The capacity and location of the charging infrastructure also play an important role in the operations of E-AMoD systems, and will influence the rebalancing schemes 
of the vehicles. ~\cite{LukeSalazarEtAl2021} proposed a model to jointly optimize the operations and charging infrastructure for an E-AMoD system which, however, requires a long time on high performance hardware to find the solution.
Yet having a tractable problem can be extremely important, especially when the goal is to explore different parametrizations or perform comparison studies, e.g., in terms of vehicular composition of the fleet as is the case in this paper.

Few papers have focused on the joint design and optimization of a fleet for AMoD.
\cite{Wallar_2019} investigated multi-class fleet composition for shared mobility-as-a-service, while  \cite{PaparellaHofmanEtAl2022} leveraged directed acyclic graphs to study the trade-off between number of vehicles, battery capacity and costs of operations in a fleet for AMoD. However, the majority of these works suffer from scalability issues.

In conclusion, to the best of the authors' knowledge, there are no scalable models available to optimize the operations of an E-AMoD fleet jointly with the charging infrastructure siting in a computationally-tractable manner and with global optimality guarantees.

\textit{Statement of Contributions:} This paper presents a modeling and optimization framework to jointly optimize the operations of an E-AMoD fleet jointly with the charging infrastructure placement in a computationally-effective manner and with global optimality guarantees. 
We first propose a network flow model describing the fleet routing and charging activities combined with the infrastructure design. 
We perform a sampling of the road network to account for the battery state of charge (SoC) of the vehicles without significantly increasing computational complexity. 
Next, we frame the optimization problem in a (mixed-integer) linear fashion that can be efficiently solved with off-the-shelf algorithms in a few minutes. 
Finally, we showcase our framework with two case studies. The first case study shows the impact of the siting and density of the charging infrastructure on the energy consumed by the user-free flow of vehicles, and the second explores the impact of a given vehicle on the fleet sizing and energy consumption. Both case studies are carried out for the area of Manhattan, New York City.

\textit{Organization:} The remainder of this paper is structured as follows: Section \ref{sec: Methodology} introduces the E-AMoD optimization framework. 
Section \ref{sec: Results} details our case studies of Manhattan.
Finally, Section \ref{sec: Conclusions} draws the conclusions from our key findings and provides an outlook on future research.
\section{Methodology}\label{sec: Methodology}
In this section, we present a time-invariant network flow model to optimize the operation of an E-AMoD system jointly with the placement of the charging infrastructure. 
To this end, we construct the multi-layer digraph shown in Fig.~\ref{fig:Overview}, to represent the position of the vehicles on the road network together with their SoC.

\subsection{Multi-Layer Sampled Graph} 
\begin{figure}[t]
\begin{minipage}{0.95\linewidth}
\end{minipage}
\begin{minipage}{\columnwidth}
\includegraphics[width=8.7cm]{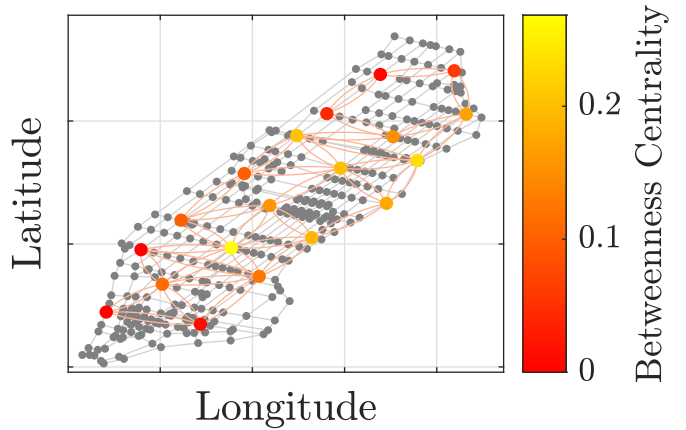}
\end{minipage}
\caption{Road graph of Manhattan (grey). Reduced iso-energy graph (colored). Each orange arc has a weight equal to the unit energy or an integer multiple.}
\label{fig:FinalGraphs}
\label{fig:Centrality}
\end{figure}

We model the transportation network as a directed graph $\mathcal{G_\mathrm{R}} = (\mathcal{V_\mathrm{R}}, \mathcal{A_\mathrm{R}})$ with a set of vertices ${v} \in \mathcal{V_\mathrm{R}}$ representing the location of intersections on the road network, and  a set of arcs $(i,j) \in \mathcal{A_\mathrm{R}}$ representing the road link between vertices $i$ and $j$. Each road arc $(i,j) \in \mathcal{A_\mathrm{R}}$ is characterized by a distance $d_{ij}$, travel time $t_{ij}$, and energy  $e_{ij}$ required to traverse it.
We then filter the original network of the city, obtaining a new network where each pair of connected nodes is equally distant in energy consumption from eachother. The selected reduced set of nodes is used to search for node pairs, which are also multiples of the unit energy consumption. This allows for increased accuracy in the path planning and the subsequent energy consumption with the reduced set of nodes.  
We highlight that the energy consumption (i.e., the unit energy of the arcs) depends on the vehicle design.
To this end, we devise an algorithmic procedure to reduce the original road network into a reduced network.
The final transportation network graph consists of a smaller set of vertices and the corresponding set of arcs between them, as for instance shown in Fig.~\ref{fig:FinalGraphs} for Manhattan, NYC. 
To model the SoC of the fleet, we build a multi-layer graph, where each layer is a copy of the previously mentioned reduced graph. The total number of layers is equal to the battery capacity of the vehicles divided by the energy discretization of the road network. 
The top layer represents the battery at $100$\%, and the bottom layer at $0$\% SoC. Therefore, as the vehicle commutes each arc, it also traverses down one layer, representing the depleting SoC.
Thanks to the new equi-distant representation in energy of the network, we eliminate the problem of dealing with the discretization of the SOC. Without this important stage, the discretization would lead to either 1) a very high number of layers and intractability, in case of fine SOC discretization; 2) uncertainty and mismatch between real energy used and the represented SoC on the layer, in case of coarse SoC discretization. 

\subsection{Travel Requests and Charging Stations}
We introduce a set of geo-nodes $\mathcal{V_\mathrm{G}}$, where the travel requests are initialized.
A geo-node is connected to all the vertices at the same geographical location, across all the battery SoC layers via geo-arcs. 
Then, we define $\mathcal{M} = \{1, ..., M\}$ as the set of travel requests. 
Each request $m \in \mathcal{M}$ is defined by a tuple ${r}_m = (o_m, d_m, \alpha_m) \in \mathcal{V_\mathrm{G}} \times \mathcal{V_\mathrm{G}} \times \mathbb{R}^+$ in which $\alpha_m$ is the number of users traveling from the origin $o_m$ to the destination $d_m$ per unit time.
The user flow induced by each demand $m$ is defined as $x_{ij}^{m}$, where $ m \in \mathcal{M}$ on the arcs $(i, j) \in  \mathcal{A}$.
All user demand flows $x_{ij}^{m}$ are fulfilled by vehicles, and the user-free vehicle flows are defined as $x_{ij}^{\mathrm{r}}$, which is the rebalancing flow on arc $(i,j) \in \mathcal{A}$. Both the user demand flow and the rebalancing flow originate and terminate at the set of geo-nodes $\mathcal{V_\mathrm{G}}$. 

We also include the presence of charging stations that are defined by the binary variable $c_i \in \mathcal{V_\mathrm{G}}$, where the set of geo-nodes $\mathcal{V_\mathrm{G}}$ is where they can be located (we recall that a geo-node connects all nodes in the graph that represents the same geographical location). If $c_i=1$, node $i$ is a charging station, if $c_i=0$, there is no charging station in node $i$.
Charging arcs $(i,j) \in \mathcal{A}_\mathrm{C}$ are directed arcs at charging station locations, which allow the vehicles to move from a layer to the subsequent layer above,  while remaining at the same geographic location. Vehicles can only be charged while rebalancing via charging arcs $(i,j) \in \mathcal{A_\mathrm{C}}$.
Hence, the set of arcs $\mathcal{A_\mathrm{R}} \cup \mathcal{A_\mathrm{C}} \cup \mathcal{A_\mathrm{G}} \in \mathcal{A}$, and the set of vertices  $\mathcal{V_\mathrm{R}} \cup \mathcal{V_\mathrm{G}} \in \mathcal{V}$ comprise the complete graph representation.

\subsection{Problem Formulation}
The objective of this paper is to minimize the user and rebalancing flows in the system:
\begin{equation}\label{eq: objective1}
    \min_{{x}^{m},{x}^\mathrm{r}}\sum_{{m} \in \mathcal{M}} \sum_{{(i,j)}\in\mathcal{A}} t_{ij}\cdot (x_{ij}^{m} + x_{ij}^\mathrm{r}),
\end{equation}
where $x_{ij}^{m}$ is the user demand flow, $x_{ij}^\mathrm{r}$ is the rebalancing flow, $t_{ij}$ is the time to traverse arc ${(i,j)}\in\mathcal{A}$.
The vehicle and user flow conservation are expressed as in a multi-commodity transportation problem by
\begin{multline}\label{eq: Flow conservation}
        \sum_{(i,j)\in\mathcal{A}}x_{ij}^{{m}} + \mathds{1}_{j=o_{m}}\cdot \alpha_{{m}} = \sum_{(j,k)\in\mathcal{A}}x_{jk}^{{m}}+ \mathds{1}_{j=d_{m}}\cdot \alpha_{{m}}\\ \qquad \forall {m}\in\mathcal{M},
\end{multline}
where the user flow $x_{ij}^{m}$ is induced by demand $m$, $\mathds{1}_{x=y}$ is the indicator function, equal to $1$ if $x=y$ and zero otherwise, and $\alpha_m$ is the user request rate per unit time.
Rebalancing the vehicles in the E-AMoD system is critical to create a balanced system and to re-align vehicle distribution with transportation requests. This is ensured by
\begin{equation} \label{eq: Flow Rebalance}
        \sum_{(i,j)\in\mathcal{A}} \left( x_{ij}^{\mathrm{r}} + \sum_{{m} \in \mathcal{M}} x_{ij}^{m}   \right) = 
         \sum_{(j,k)\in\mathcal{A}} \left( x_{jk}^{\mathrm{r}} + \sum_{{m} \in \mathcal{M}} x_{jk}^{m}   \right).
\end{equation}
The geo-nodes $\mathcal{V}_\mathrm{G}$ act as origins and destinations for the vehicles. Consequently, each demand will require the geo-arcs $\mathcal{A}_\mathrm{G}$ to be used twice: first, from the origin geo-node to the road network, and second, to go from the road network to the destination geo-node $\mathcal{V}_\mathrm{G}$. The same is also applicable to rebalancing flows. This is ensured by constraining the users flow by
\begin{equation}\label{eq: FromGeo_vehicle}
    \begin{aligned}
        \sum_{{m \in \mathcal{M}}} \left(\sum_{(i,j) \in \mathcal{A}_\mathrm{G}} x_{ij}^{m}+\sum_{(k,l) \in \mathcal{A}_\mathrm{G}} x_{kl}^{m} \right) = 2 \cdot \sum_{{m}\in\mathcal{M}} \alpha_{{m}} \\ \qquad \forall {i,l} \in \mathcal{V}_\mathrm{G},
    \end{aligned}
\end{equation}
and the rebalancing flows to
\begin{equation}\label{eq: FromGeo_rebalance}
    \begin{aligned}
        \sum_{(i,j) \in \mathcal{A}_\mathrm{G}} x_{ij}^\mathrm{r}+\sum_{(k,l) \in \mathcal{A}_\mathrm{G}} x_{kl}^\mathrm{r} = 2 \cdot \sum_{{m}\in\mathcal{M}} \alpha_{{m}}  \qquad \forall {i,l} \in \mathcal{V}_\mathrm{G}.
    \end{aligned}
\end{equation}
We enforce SoC conservation when passing through a geo-node $\mathcal{V}_{\mathrm{G}}$ as
\begin{equation}\label{eq: vehicleRebalance_Geo}
    \begin{aligned}
        \sum_{(i,j) \in \mathcal{A}_\mathrm{G}} x_{ij}^{m} - x_{ji}^\mathrm{r} = 0 \qquad \forall {m}\in\mathcal{M}, \forall {i,j} \in \mathcal{V}_\mathrm{G}.
    \end{aligned}
\end{equation}
We limit the number of charging stations to $N\in\sN$ with
\begin{equation}\label{eq: sumCP}
    \begin{aligned}
        \sum_{i \in  \mathcal{V}_\mathrm{G}} c_{i} \leq N.
    \end{aligned}
\end{equation}
Each charging station has a limited capacity in terms of the number of vehicles it can charge simultaneously. The capacity constraint of each charging station is given by
\begin{equation}\label{eq: chargingArcs}
    \begin{aligned}
    x_{ij}^{\mathrm{r}} \leq  Z \cdot E \qquad \forall {(i,j)} \in \mathcal{A_\mathrm{C}},
    \end{aligned}
\end{equation}
where $Z$ refers to the maximum vehicle capacity of each charging station, and $E$ refers to the number of battery SoC layers that can be traversed via the charging arcs per unit time, i.e., the charging power at the charging stations. 
Therefore, the charging station vehicle capacity and the charging power limit the rebalancing flow through the charging arcs ${(i,j)} \in \mathcal{A_\mathrm{C}}$. 
Since the vehicles must not charge with users onboard, the charging of the vehicles can be carried out only while rebalancing. This is ensured by
\begin{equation}\label{eq: Xv(charging) = 0}
    \begin{aligned}
        x_{ij}^{{m}} = 0 \qquad \forall {m}\in\mathcal{M}, \forall {(i,j)} \in \mathcal{A}_\mathrm{C},
    \end{aligned}
\end{equation}
which guarantees that the vehicles do not charge when catering to user demands. Therefore, the vehicles can only charge while rebalancing through charging arcs $(i,j) \in \mathcal{A_\mathrm{C}}$.
Finally, we impose non-negative flow constraints,
\begin{align}
&x_{ij}^\mathrm{{r}} \geq 0 \qquad \forall {(i,j)} \in \mathcal{A}, \label{eq: Xv > 0} \\
&x_{ij}^{{m}} \geq 0 \qquad \forall {m}\in\mathcal{M}, \forall {(i,j)} \in \mathcal{A}.  \label{eq: Xu > 0}
\end{align}

First, given a pre-defined charging infrastructure placement, we define the E-AMoD optimization problem as follows:
\begin{prob}(E-AMoD Optimization Problem:)\label{prob:one}
Given a set of transportation requests $\mathcal{M}$, the optimal user flows $x^m$, and rebalancing flows $x^r$ result from:
\begin{equation*}
\begin{aligned}
&\!\min_{{x}^{m},{x}^\mathrm{r}}\sum_{{m} \in \mathcal{M}} \sum_{{(i,j)}\in\mathcal{A}} & & t_{ij}\cdot (x_{ij}^{m} + x_{ij}^\mathrm{r}), \\
& \textnormal{s.t. } & &\eqref{eq: Flow conservation}-\eqref{eq: vehicleRebalance_Geo}, \eqref{eq: Xv(charging) = 0}-\eqref{eq: Xu > 0} .
\end{aligned}
\end{equation*}
\end{prob}

Problem~\ref{prob:one} is a linear program (LP) that can be efficiently solved with global optimality guarantees with off-the-shelf LP solvers.

\begin{prob}(E-AMoD Joint Optimization Problem:)\label{prob:two}
Given a set of transportation requests $\mathcal{M}$, the optimal user flows $x^m$, and rebalancing flows $x^r$ and the optimal siting of the charging infrastructure $c$, result from:
\begin{equation*}
\begin{aligned}
&\!\min_{{x}^{m},{x}^\mathrm{r},c}\sum_{{m} \in \mathcal{M}} \sum_{{(i,j)}\in\mathcal{A}} & & t_{ij}\cdot (x_{ij}^{m} + x_{ij}^\mathrm{r}), \\
& \textnormal{s.t. } & &\eqref{eq: Flow conservation}-\eqref{eq: Xu > 0} .
\end{aligned}
\end{equation*}
\end{prob}

Problem~\ref{prob:two} is a mixed-integer linear program (MILP) that can be solved with global optimality guarantees with off-the-shelf MILP solvers.

\subsection{Discussion}
A few comments are in order. 
First, we model the system at steady-state. This assumption holds if the rate change of requests is significantly lower than the average travel time of individual trips, as observed in densely populated urban environments by \cite{Neuburger1971, Rossi2018}. In addition, vehicle conservation implicitly enforce SoC conservation over the time-span under consideration.
Second, the iso-energy graph sampling approach is only valid for environments where traveling to and from a node requires the same amount of energy, e.g., flat urban environments, and should be extended to capture more general (e.g., hilly) scenarios as well.
Third, the results obtained by solving Problem \ref{prob:one} and \ref{prob:two} allow for fractional flows to occur. This is acceptable, given the mesoscopic nature of the problem, where arc flows are in the order of hundreds of vehicles, see~\cite{LukeSalazarEtAl2021}.
Fourth, we do take into account the exogenous traffic and its stochatiscity only for a specific traffic scenario. In case of different conditions,  the travel time and energy consumption should be updated, and consequently resample the network. Finally, we do not take into account the impact of endogenous traffic, e.g., as done in~\cite{SalazarTsaoEtAl2019,Wollenstein-BetechSalazarEtAl2021}, but rather leave this interesting aspect to future research.

\section{Results} \label{sec: Results}
This section showcases our modeling and optimization framework in two real-world case studies for Manhattan, NYC. The original data set is extracted from OpenStreetMap \citep{HaklayWeber2008}, and reduced to 19 nodes and 138 arcs, as shown in Fig.~\ref{fig:FinalGraphs}.
The transportation demand requests are available publicly (courtesy of the New York Taxi and Limousine Commission).
The data set used as a reference is taken from March 1 to 10, 2022. Thereby, we consider approximately \unit[140]{thousand} demands per day and \unit[1.4]{million} demands for the entire 10-day period. Due to the high volume and the absence of strong peak hours of daily travel requests in NYC, see ~\cite{Meyers2018}, we do not lose the assumptions of the linear time-invariant model. Thus, we can use it to model a time period as long as multiple days. 
Moreover, the results that will be shown in Section~\ref{sec: caseStudy1}, specifically the ratio between rebalancing and overall distance driven, are in line with~\cite{HogeveenSteinbuchEtAl2021}, enforcing that the linear time-invariant hypothesis over a day holds.

We investigate two case studies. The first one assesses the advantages of jointly optimizing the infrastructure siting with respect to a heuristic placement based on geo-nodes centrality, and the impact of the charging infrastructure density on the resulting rebalancing energy consumed in a day by the whole fleet.
The second case study evaluates the performance of the E-AMoD system when different types of electric vehicles are employed, see Table~\ref{table:Vehicle}, which have differently sized batteries and accordingly designed powertrains.
\begin{table}[t]
\begin{center}
\caption{Normalized Vehicle Parameters and Number of Layers in the Graph.}
\label{table:Vehicle}
\begin{tabular}{ l|l|l|l }
 
  & Car A & Car B & Car C  \\ 
  \hline
  Energy consumption (WLTP) & 85\% & 93\% & 100\%  \\
  Battery capacity & 25\% & 60\% & 100\%  \\
  Mass & 69\% & 85\% & 100\% \\
  Number of SoC Layers & 196  & 428 & 670
\end{tabular}
\end{center}
\end{table}
Car A has the smallest battery capacity, and is therefore the lightest and hence most energy-efficient vehicle.
Conversely, Car C has the largest battery capacity, making it the heaviest and least efficient one.
For all case-studies, both Problem~\ref{prob:one} and Problem~\ref{prob:two} were solved using the solver Gurobi 9.5, see \cite{GurobiOptimization2021}, on an Intel core i7-10850H, 32GB RAM in less than 5 and 15 minutes, respectively.

\subsection{The Charging Infrastructure}\label{sec: caseStudy2}
In this case study we investigate the impact of the charging infrastructure density and siting on the user-free flows. We simulate a demand period of 10 days with Car B. 
Each data point in the figure is the average  daily energy usage of the user-free flow between 1-10 March, 2022.
We assess the advantages of jointly optimizing siting and routing by comparing the results of Problem~\ref{prob:two} with the ones of Problem~\ref{prob:one} after siting the charging infrastructure in a heuristic manner. Thereby, we adopt a heuristic policy that places the charging stations in the geo-nodes with the highest \textit{betweenness} centrality,~\cite{Bullo2018},---i.e., the probability that a node appears on the shortest path between any two random nodes, ---as shown in Fig.~\ref{fig:Centrality}.
\begin{figure}[t]
	\centering
	\includegraphics[width=\columnwidth]{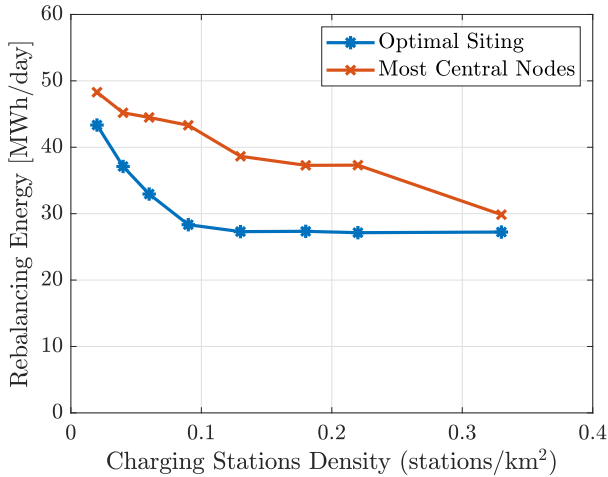}
	\caption{The average energy consumption of the user-free flow for an optimally sited charging infrastructure and a heuristic approach based on betweenness centrality. Data set from 1-10 March, 2022.}
	\label{fig:Rebal}
\end{figure}

Fig.~\ref{fig:Rebal} shows the difference in the energy usage of the user-free flows in the two scenarios for an increasing number of charging stations.
The optimal siting significantly outperforms the heuristic approach, with up to 30\% decrease in energy consumption.
Moreover, we observe that with approximately $\unit[0.1]{stations/km^2}$ , if the siting is optimal, we reach a plateau up to which it is not convenient to build additional stations. On the contrary, by using the betweenness centrality heuristic method, the number of charging stations has to be at least 3 times higher to obtain a similar performance.

\subsection{Impact of the Vehicle Design on the Optimal Operations}\label{sec: caseStudy1}
This section details the case study for the impact of selected vehicles on the sizing and energy consumption of the fleet. Specifically, we select 3 different vehicles that are used to study the impact on the comparative metrics. The design parameters for these vehicles are normalized with respect to the vehicle from the Dutch solar car manufacturer Lightyear, see \cite{Lightyear}. Table \ref{table:Vehicle} shows the normalized parameters of the selected vehicles and the number of layers used in the multi-layer graph. The simulation is carried out for the whole day of March 1, 2022. In each scenario we solve Problem~\ref{prob:two} with $N=10$ charging stations, following the results of Section~\ref{sec: caseStudy2}.
\begin{figure}[t]
    \centering
    \includegraphics[width=\columnwidth]{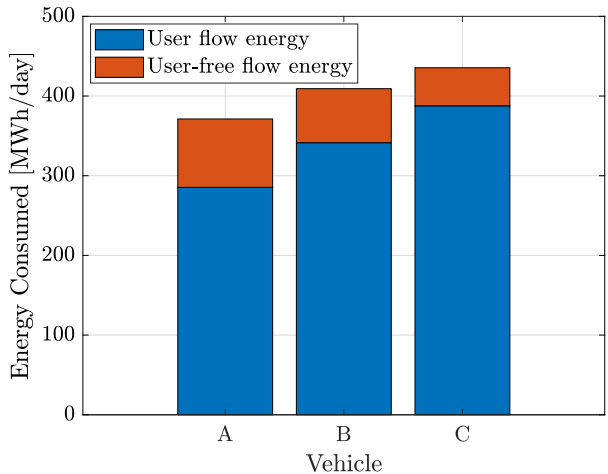}
    \caption{Energy used by user flow (blue) and user-free flow (red) for different vehicles (A,B,C) in Manhattan. The resulting fleet sizes are equal to 7910, 7890 and 7880 vehicles, respectively.}
    \label{fig:CS1}
\end{figure}
Fig.~\ref{fig:CS1} illustrates the overall energy consumed and the one exclusively caused by the rebalancing.
The energy consumption of Car A is about $15$\% less than Car C during a WLTP. 
However, a vehicle with a larger battery requires fewer trips to the charging station, as reflected in Fig.~\ref{fig:CS1}. This results in a higher energy consumption of the rebalancing flow for Car A. 
Nevertheless, the overall energy consumption of Car A is still the lowest due to its significantly higher energy efficiency resulting from a lower weight. 
In particular, we see that Car A overall uses $12$\% less energy compared to Car C, even though it uses $44$\% more energy to rebalance.
Finally, the resulting fleet sizes are equal to $7910$, $7890$ and $7880$ vehicles for A,B and C, respectively. With these results, the difference in fleet size can be considered negligible with respect to the battery size of the individual vehicle.  
In conclusion, our results indicate that using vehicles with a shorter driving range but a higher energy efficiency can lead to better performance because the possibility of a better charging trip scheduling is overcompensated by a lower total energy need.

\section{Conclusion}\label{sec: Conclusions}
This paper proposed a modeling and optimization framework to jointly optimize operations and charging infrastructure placement for Electric Autonomous Mobility-on-Demand (E-AMoD) systems in a computationally tractable, scalable and globally optimal fashion.
The proposed iso-energy graph resampling allowed us to construct a time-invariant network flow model that can account for the State of Charge (SoC) of the individual vehicles, without the actual fleet size affecting its computational effectiveness.
Leveraging this model, we formulated the E-AMoD operational-only and joint operation and infrastructure placement problems as a linear program (LP) and mixed-integer linear program (MILP), respectively, that could be solved with global optimality guarantees using off-the-shelf optimization algorithms in a few minutes only.
Our real-world case studies based on taxi data in Manhattan, NYC, showed that jointly optimizing the charging infrastructure placement can significantly improve the performance achievable by the E-AMoD system. It also revealed that the benefit of increasing the number of charging stations vanishes rapidly in case of optimal siting of the infrastructure, but this might not occur if the placement is heuristic.
Moreover, in line with our previous findings obtained in~\cite{PaparellaHofmanEtAl2022} with a completely different, still globally optimal method, we showed how deploying fleets with a  downsized battery will slightly increase the empty-mileage driven, whilst resulting in almost the same fleet size and significantly reducing the total energy consumption.

This study opens the field for the following extensions:
First, we would like to include the total costs of ownership of the fleet and charging infrastructure, thereby also accounting for different levels of charging power.
Second, we deem it interesting to also capture the interactions of the fleet with public transit and (shared) active modes.
Finally, it would be insightful to account for the interactions with the power grid and study heterogeneous (solar) electric fleet compositions.

\section{Acknowledgments}\label{Sec:akn}
We thank Dr. I. New and Ir. O. Borsboom for proofreading this paper.

                                                   






\bibstyle{ifacconf}
\bibliography{main.bib,SML_papers.bib}

\newcommand{\noopsort}[1]{} \newcommand{\printfirst}[2]{#1}
  \newcommand{\singleletter}[1]{#1} \newcommand{\switchargs}[2]{#2#1}
\begin{thebibliography}{26}
\providecommand{\natexlab}[1]{#1}
\providecommand{\url}[1]{\texttt{#1}}
\providecommand{\urlprefix}{URL }
\expandafter\ifx\csname urlstyle\endcsname\relax
  \providecommand{\doi}[1]{doi:\discretionary{}{}{}#1}\else
  \providecommand{\doi}{doi:\discretionary{}{}{}\begingroup
  \urlstyle{rm}\Url}\fi

\bibitem[{Banerjee et~al.(2015)Banerjee, Johari, and
  Riquelme}]{BanerjeeJohariEtAl2015}
Banerjee, S., Johari, R., and Riquelme, C. (2015).
\newblock Pricing in ride-sharing platforms: A queueing-theoretic approach.
\newblock In \emph{{ACM Conf.\ on Economics and Computation}}.

\bibitem[{Boewing et~al.(2020)Boewing, Schiffer, Salazar, and
  Pavone}]{BoewingSchifferEtAl2020}
Boewing, F., Schiffer, M., Salazar, M., and Pavone, M. (2020).
\newblock A vehicle coordination and charge scheduling algorithm for electric
  autonomous mobility-on-demand systems.
\newblock In \emph{{American Control Conference}}.

\bibitem[{Bullo(2020)}]{Bullo2018}
Bullo, F. (2020).
\newblock \emph{Lectures on Network Systems}.
\newblock Kindle Direct Publishing, 1.4 edition.
\newblock \urlprefix\url{http://motion.me.ucsb.edu/book-lns}.
\newblock With contributions by J. Cortes, F. Dorfler, and S. Martinez.

\bibitem[{Estandia et~al.(2019)Estandia, Schiffer, Rossi, Kara, Rajagopal, and
  Pavone}]{EstandiaSchifferEtAl2019}
Estandia, A., Schiffer, M., Rossi, F., Kara, E.C., Rajagopal, R., and Pavone,
  M. (2019).
\newblock On the interaction between autonomous mobility on demand systems and
  power distribution networks -- an optimal power flow approach.
\newblock \emph{arXiv preprint arXiv:1905.00200}.
\newblock \urlprefix\url{https://arxiv.org/abs/1905.00200}.

\bibitem[{{Gurobi Optimization, LLC}(2021)}]{GurobiOptimization2021}
{Gurobi Optimization, LLC} (2021).
\newblock Gurobi optimizer reference manual.
\newblock {Available at }\url{http://www.gurobi.com}.

\bibitem[{Haklay and Weber(2008)}]{HaklayWeber2008}
Haklay, M. and Weber, P. (2008).
\newblock {OpenStreetMap}: User-generated street maps.
\newblock \emph{{IEEE Pervasive Computing}}, 7(4), 12--18.

\bibitem[{Hogeveen et~al.(2021)Hogeveen, Steinbuch, Verbong, and
  Hoekstra}]{HogeveenSteinbuchEtAl2021}
Hogeveen, P., Steinbuch, M., Verbong, G., and Hoekstra, A. (2021).
\newblock Quantifying the fleet composition at full adoption of shared
  autonomous electric vehicles: An agent-based approach.
\newblock \emph{The Open Transportation Journal}, 15, 47--60.

\bibitem[{H{\"o}rl et~al.(2019)H{\"o}rl, Ruch, Becker, Frazzoli, and
  Axhausen}]{HorlRuchEtAl2019}
H{\"o}rl, S., Ruch, C., Becker, F., Frazzoli, E., and Axhausen, K. (2019).
\newblock Fleet operational policies for automated mobility: A simulation
  assessment for zurich.
\newblock \emph{{Transportation Research Part C: Emerging Technologies}}, 102,
  20--31.
\newblock \doi{https://doi.org/10.1016/j.trc.2019.02.020}.

\bibitem[{Iglesias et~al.(2018)Iglesias, Rossi, Wang, Hallac, Leskovec, and
  Pavone}]{IglesiasRossiEtAl2018}
Iglesias, R., Rossi, F., Wang, K., Hallac, D., Leskovec, J., and Pavone, M.
  (2018).
\newblock Data-driven model predictive control of autonomous mobility-on-demand
  systems.
\newblock In \emph{{Proc.\ IEEE Conf.\ on Robotics and Automation}}.

\bibitem[{Iglesias et~al.(2019)Iglesias, Rossi, Zhang, and
  Pavone}]{IglesiasRossiEtAl2017}
Iglesias, R., Rossi, F., Zhang, R., and Pavone, M. (2019).
\newblock A {BCMP} network approach to modeling and controlling autonomous
  mobility-on-demand systems.
\newblock \emph{{Proc.\ of the Inst.\ of Mechanical Engineers, Part~D: Journal
  of Automobile Engineering}}, 38(2--3), 357--374.

\bibitem[{Levin et~al.(2017)Levin, Kockelman, Boyles, and
  Li}]{LevinKockelmanEtAl2017}
Levin, M.W., Kockelman, K.M., Boyles, S.D., and Li, T. (2017).
\newblock A general framework for modeling shared autonomous vehicles with
  dynamic network-loading and dynamic ride-sharing application.
\newblock \emph{Computers, Environment and Urban Systems}, 64, 373 -- 383.

\bibitem[{Lightyear(2016)}]{Lightyear}
Lightyear (2016).
\newblock Available online at \url{https://lightyear.one/} Accessed:
  26/09/2022.

\bibitem[{Luke et~al.(2021)Luke, Salazar, Rajagopal, and
  Pavone}]{LukeSalazarEtAl2021}
Luke, J., Salazar, M., Rajagopal, R., and Pavone, M. (2021).
\newblock Joint optimization of electric vehicle fleet operations and charging
  station siting.
\newblock In \emph{{Proc.\ IEEE Int.\ Conf.\ on Intelligent Transportation
  Systems}}.
\newblock In press.

\bibitem[{Maciejewski et~al.(2017)Maciejewski, Bischoff, H\"orl, and
  Nagel}]{MaciejewskiBischoffEtAl2017}
Maciejewski, M., Bischoff, J., H\"orl, S., and Nagel, K. (2017).
\newblock Towards a testbed for dynamic vehicle routing algorithms.
\newblock In \emph{{Int.\ Conf.\ on Practical Applications of Agents and
  Multi-Agent Systems - Workshop on the application of agents to passenger
  transport (PAAMS-TAAPS)}}.

\bibitem[{Meyers(2018)}]{Meyers2018}
Meyers, A. (2018).
\newblock Analyzing over 450 million taxi trips using hadoop and pyspark.
\newblock Available Online at :
  https://www.linkedin.com/pulse/where-ya-headed-analyzing-over-450-million-taxi-trips-adrian-meyers-1/.

\bibitem[{Neuburger(1971)}]{Neuburger1971}
Neuburger, H. (1971).
\newblock The economics of heavily congested roads.
\newblock \emph{{Transportation Research}}, 5(4), 283--293.

\bibitem[{Paparella et~al.(2022)Paparella, Hofman, and
  Salazar}]{PaparellaHofmanEtAl2022}
Paparella, F., Hofman, T., and Salazar, M. (2022).
\newblock Joint optimization of number of vehicles, battery capacity and
  operations of an electric autonomous mobility-on-demand fleet.
\newblock In \emph{{Proc.\ IEEE Conf.\ on Decision and Control}}.

\bibitem[{Rossi(2018)}]{Rossi2018}
Rossi, F. (2018).
\newblock \emph{On the Interaction between {Autonomous Mobility-on-Demand}
  Systems and the Built Environment: Models and Large Scale Coordination
  Algorithms}.
\newblock Ph.D. thesis, {Stanford University, Dept.\ of Aeronautics and
  Astronautics}.

\bibitem[{Rossi et~al.(2020)Rossi, Iglesias, Alizadeh, and
  Pavone}]{RossiIglesiasEtAl2018b}
Rossi, F., Iglesias, R., Alizadeh, M., and Pavone, M. (2020).
\newblock On the interaction between {Autonomous Mobility-on-Demand} systems
  and the power network: Models and coordination algorithms.
\newblock \emph{{IEEE Transactions on Control of Network Systems}}, 7(1),
  384--397.

\bibitem[{Rossi et~al.(2018)Rossi, Zhang, Hindy, and
  Pavone}]{RossiZhangEtAl2017}
Rossi, F., Zhang, R., Hindy, Y., and Pavone, M. (2018).
\newblock Routing autonomous vehicles in congested transportation networks:
  Structural properties and coordination algorithms.
\newblock \emph{{Autonomous Robots}}, 42(7), 1427--1442.

\bibitem[{Salazar et~al.(2020)Salazar, Lanzetti, Rossi, Schiffer, and
  Pavone}]{SalazarLanzettiEtAl2019}
Salazar, M., Lanzetti, N., Rossi, F., Schiffer, M., and Pavone, M. (2020).
\newblock Intermodal autonomous mobility-on-demand.
\newblock \emph{{IEEE Transactions on Intelligent Transportation Systems}},
  21(9), 3946--3960.

\bibitem[{Salazar et~al.(2019)Salazar, Tsao, Aguiar, Schiffer, and
  Pavone}]{SalazarTsaoEtAl2019}
Salazar, M., Tsao, M., Aguiar, I., Schiffer, M., and Pavone, M. (2019).
\newblock A congestion-aware routing scheme for autonomous mobility-on-demand
  systems.
\newblock In \emph{{European Control Conference}}.

\bibitem[{Spieser et~al.(2014)Spieser, Treleaven, Zhang, Frazzoli, Morton, and
  Pavone}]{SpieserTreleavenEtAl2014}
Spieser, K., Treleaven, K., Zhang, R., Frazzoli, E., Morton, D., and Pavone, M.
  (2014).
\newblock Toward a systematic approach to the design and evaluation of
  {Autonomous} {Mobility-on-Demand} systems: A case study in {Singapore}.
\newblock In \emph{Road Vehicle Automation}. {Springer}.

\bibitem[{Wallar et~al.(2019)Wallar, Schwarting, Alonso-Mora, and
  Rus}]{Wallar_2019}
Wallar, A., Schwarting, W., Alonso-Mora, J., and Rus, D. (2019).
\newblock Optimizing multi-class fleet compositions for shared
  mobility-as-a-service.
\newblock In \emph{{Proc.\ IEEE Int.\ Conf.\ on Intelligent Transportation
  Systems}}, 2998--3005. {IEEE}.
\newblock \doi{10.1109/itsc.2019.8916904}.

\bibitem[{Wollenstein-Betech et~al.(2021)Wollenstein-Betech, Salazar,
  Houshmand, Pavone, Cassandras, and
  Paschalidis}]{Wollenstein-BetechSalazarEtAl2021}
Wollenstein-Betech, S., Salazar, M., Houshmand, A., Pavone, M., Cassandras,
  C.G., and Paschalidis, I.C. (2021).
\newblock Routing and rebalancing intermodal autonomous mobility-on-demand
  systems in mixed traffic.
\newblock \emph{{IEEE Transactions on Intelligent Transportation Systems}}.
\newblock In press.

\bibitem[{Zhang and Pavone(2016)}]{ZhangPavone2016}
Zhang, R. and Pavone, M. (2016).
\newblock Control of robotic {Mobility-on-Demand} systems: A
  queueing-theoretical perspective.
\newblock \emph{{Proc.\ of the Inst.\ of Mechanical Engineers, Part~D: Journal
  of Automobile Engineering}}, 35(1--3), 186--203.

\end{thebibliography}

\end{document}